\documentclass[a4paper]{jpconf}

\usepackage{amsmath,amssymb,epsfig}
\usepackage{wrapfig}
\usepackage{gensymb}
\usepackage[english]{babel}
\usepackage{graphics}
\usepackage{subcaption}
\usepackage{hyperref}
\usepackage[utf8]{inputenc}

\usepackage{booktabs}

\newcommand{\be}{\begin{equation}}
\newcommand{\ee}{\end{equation}}

\begin{document}
\title{Summing up Ultra-High-Energy Cosmic Rays from Radio Galaxies}

\author{Bj\"{o}rn Eichmann}

\address{Ruhr Astroparticle and Plasma Physics Center (RAPP Center), Ruhr-Universit\"at Bochum, Institut f\"ur 
Theoretische Physik IV/ Plasma-Astroteilchenphysik, 44780 Bochum, Germany}

\ead{eiche@tp4.rub.de}

\begin{abstract}
Radio galaxies are intensively discussed as the sources of cosmic rays observed above about $3\,\text{EeV}$, called ultra-high-energy cosmic rays (UHECRs). 
Here, the key issues from a recent investigation are summed up, where the individual characteristics of radio galaxies, as well as the impact by the extragalactic magnetic-field structures up to a distance of 120\,\text{Mpc} has been taken into account. 
It is shown that the average contribution of radio galaxies taken over a very large volume cannot explain the observed features of UHECRs measured at Earth. 
However, we obtain excellent agreement with the spectrum, composition, and arrival-direction distribution of UHECRs measured by the Pierre Auger Observatory, if we assume that most UHECRs observed arise from only two sources: The ultra-luminous radio galaxy Cygnus A, providing a mostly light composition of nuclear species dominating up to about $60\,$EeV, and the nearest radio galaxy Centaurus A, providing a heavy composition dominating above $60\,$EeV. Here we have to assume that extragalactic magnetic fields out to $250\,$Mpc, which we did not include in the simulation, are able to isotropize the UHECR events at about $8\,$EeV arriving from Cygnus A. 
\end{abstract}

\section{Introduction}
The origin of the Ultra-High-Energy Cosmic Rays (UHECRs) is still one of the great enigmas of modern astrophysics.
Basically, there are three main observational characteristics, from observatories like the Pierre Auger Observatory (Auger) in the southern hemisphere, and the High Resolution Fly's eye (HiRes) as well as the Telescope Array (TA) experiments in the northern hemisphere, that describe our current knowledge of the UHECRs:
\begin{enumerate}
 \item[(1.)] The energy spectrum, which changes above $\sim 3\,\text{EeV}$ --- the so-called ankle --- to a flatter power-law distribution with a spectral index of 2.6 and a sharp flux suppression above about $10^{19.5}\,\text{eV}$ \cite{Abbasi:2007sv, 2010PhLB..685..239A}. 
 \item[(2.)] The chemical composition, that shows an increase of the fraction of heavier elements above $10^{18.3}\,\text{eV}$ \cite{Abraham:2010yv,Aab:2014aea, ObservatoryMichaelUngerforthePierreAuger:2017fhr}. 
 \item[(3.)] The arrival directions, where recently Auger reported a $5\sigma$ detection of a dipole with an amplitude of $\approx 6.5\%$ in the UHECR arrival directions, while higher-order multipoles are still consistent with isotropy \cite{Aab:2017tyv}.
\end{enumerate}
According to a recent study by Eichmann et al. \cite{1475-7516-2018-02-036} --- hereafter referred to as E+18 --- one of 
the key questions in this field (still) is: Can the origin of those extremely energetic particles be explained by radio 
galaxies? 

In this paper, the extensive outcome of E+18 is summed up and discussed in the light of subsequent follow-up 
investigations. 
The UHECR propagation simulations in E+18 comprise three components: 
(a) a 3D structure of radio galaxies and EGMF within a radius of 120 Mpc; (b) a continuous source function derived from a luminosity function of radio galaxies for the contributions from beyond 120 Mpc; (c) a contribution of the powerful radio galaxy Cygnus A, which is near the magnetic horizon and is expected to deliver the dominant contribution to UHECRs due to its extreme radio power and brightness. 
All simulations are carried out with CRPropa3 \cite{1475-7516-2016-05-038}.

\section{Radio Galaxies as UHECR accelerators}
Radio galaxies and in particular their jets are suitable sites for first-order Fermi acceleration at their discontinuities, as shown in numerous works in the past. 
Since the Fermi acceleration process is only dependent on rigidity, abundances $f_i$ of different nuclear species with 
charge $Z_i$ measured at a given rigidity $R$ remain unchanged in the acceleration process. The abundances at injection, 
however, will depend on the environment or details of the injection process. 
In E+18, we suggest a simple relation where particles are injected at some minimal rigidity $\check{R}$ with abundances 
$f_i = f_{\odot} Z_i^q$, where $f_{\odot}$ denotes solar abundances, that enables to increase the heaviness of the 
composition based on a single parameter $q$. 

As shown in great detail in E+18, the established correlation between the jet power $Q_{\rm jet}$ and the extended radio luminosity $L_{1.1}$ from Willott et al.\ \cite{1999MNRAS.309.1017W} gives for the minimal energy condition \cite{1970ranp.book.....P} a good estimate of the cosmic ray luminosity 
\be
Q_{\rm cr}\simeq \frac47 Q_{\rm jet} = 1.3\times 10^{42}\,g_{\rm cr}\, \left({L_{1.1} \over L_\star} \right)^{6/7}\,\frac{\text{erg}}{\text{s}}\,,
\label{Qcr}
\ee
and the maximal rigidity
\be
\hat{R}(Q_{\rm cr}) = g_{\rm acc} \sqrt{Q_{\rm cr}/c}\,,
\label{Rmax}
\ee
where $L_\star \approx 4.9\times 10^{40}\,\text{erg/s}$ denotes a characteristic luminosity according to Mauch and Sadler \cite{2007MNRAS.375..931M}.
In doing so, we suppose that the total power in CRs is significantly higher than in relativistic electrons. 
From the uncertain efficiency of converting jet internal energy into observable radio luminosity, as well as the uncertain details of the acceleration process we obtain the dimensionless coefficients 
\be
1 \lesssim g_{\rm cr} \lesssim 50\quad{\rm and}\quad 
0.01 \lesssim g_{\rm acc} \lesssim 0.5\;.
\ee

Introducing different nuclei species $i$ with charge number $Z_i$ and an abundance $f_i$, the total cosmic ray power per charge number yields
\be
Q_{{\rm cr},i}\equiv Q_{\rm cr}(Z_i)=f_i\,Z_i\,Q_{\rm cr}/\bar Z\,.
\label{eq:totCRpowerPerZi}
\ee
\section{Simulation setup}
While the previously introduced physics are used to characterize the individual CR sources, the Monte Carlo based algorithm of CRPropa3 \cite{1475-7516-2016-05-038} is used to describe the propagation of UHECR nuclei in extragalactic background light, i.e.\ the cosmic microwave background (CMB) and the UV/optical/IR background (IRB), and cosmic magnetic fields. 
For the latter, the local EGMF structure from Dolag et al.\ \cite{2005JCAP...01..009D} --- hereafter referred to as D05 --- which extends up to a distance of 120\,Mpc as well as the GMF model of Jansson \& Farrar \cite{2012ApJ...757...14J} is adopted. 
However, an impact on the energy distribution by the GMF, i.e.\ energy losses or rigidity dependent arrival probabilities, is not taken into account. 
Due to rather technical issues like the limited EGMF structure and a limited CPU time, there is a need to distinguish the source contributions in the simulation setup.
\subsection{Local sources}
Sources up to a distance of 120 Mpc from Earth, called \emph{local} RGs, are treated as individual sources in a 3D simulations, where the impact of cosmic magnetic fields is taken into account. 
To obtain a complete sample of radio galaxies powerful enough to contribute to the UHECR spectrum we use the catalog of van Velzen et al.\ \cite{2012A&A...544A..18V} --- hereafter referred to as \emph{vV12 catalog}. 
After excluding 52 starforming galaxies, we are finally left with 121 radio galaxies within the considered volume, predominantly of FR-I type.  
Due to the flux limitation of the catalog as well as the limited sky coverage (88\%), the catalog misses a multiplicity of predominantly low luminous, rather distant RGs. 
In E+18 it is described in great detail, that the radio luminosity function (RLF) can be used to complete the sample, however, the resulting contribution of these missing sources to the total UHECR flux is shown to be negligible. 
Thus, each source isotropically emits an individual UHECR flux that is \emph{absolutely normalized} based on its CR power (\ref{Qcr}) and provides a maximal rigidity according to Eq.~(\ref{Rmax}).
\subsection{Average non-local sources}
Sources beyond a distance of 120 Mpc from Earth, called \emph{non-local} RGs, are treated by a continuous source function (CSF) in a 1D simulation, where only the impact of cosmic photon targets is taken into account. 
In doing so, the local CSF, $\Psi_{i,0}(R)$, is derived as
\begin{equation}
\Psi_{i,0}(R) \equiv {\mathrm{d}N_{\rm cr}(Z_i) \over \mathrm{d}V \mathrm{d}R\,\mathrm{d}t} = \int_{\check Q_{\rm cr}}^{\hat Q_{\rm cr}} S_i\big(R,\hat R(Q_{\rm cr})\big)\,{\mathrm{d}N_{\rm RG} \over \mathrm{d}V\,\mathrm{d}Q_{\rm cr}}\,\mathrm{d}Q_{\rm cr}\,,
\label{CRsourceRateDensity}
\end{equation}
using the local RLF, $\mathrm{d}N_{\rm RG}/(\mathrm{d}V\,\mathrm{d}Q_{\rm cr})$, and the CR spectrum, $S_i\big(R,\hat R(Q_{\rm cr})\big)$, of element species $i$ with charge number $Z_i$, emitted by a source with total cosmic ray power per charge number, $Q_{{\rm cr},i}$, up to a maximal rigidity $\hat R(Q_{\rm cr})$.
The limits of integration are the smallest, $\check Q_{\rm cr}$, respectively largest, $\hat Q_{\rm cr}$, CR powers we need to consider. 
To solve the integral analytically, as shown in E+18, a sharp cutoff of the individual source spectra at $\hat R(Q_{\rm cr})$ is supposed. 
The function $\Psi_{i,0}(R)$ is the local continuous source function as it is derived from a radio luminosity function determined in the local universe ($z<0.3$). To extend it to larger redshifts, we use the approximation
\be
\Psi_i(R,z) \approx \Psi_{i,0}(R)\,(1+z)^{m-1}\,,
\label{eq:SourceFunctionEvolution}
\ee
where $m \in [2,4]$ is called the source evolution index. 
The analytical solution of the CSF (\ref{eq:SourceFunctionEvolution}) shows a spectral break at a critical rigidity $R_* 
= g_{\rm acc} \sqrt{g_{\rm cr}\,Q_* / c} \approx 2\times 10^{18}\,g_{\rm acc}\,\sqrt{g_{\rm cr}}\,\text{V}$, with 
$\Psi_i(R<R_*,z)\propto R^{-a}$, where $a$ denotes the spectral index of the individual sources, and 
$\Psi_i(R>R_*,z)\propto R^{-3}$ or even steeper.
This means that for light nuclei the spectral index in the UHECR regime is too steep to explain the observed UHECR 
spectrum (as propagation will make the spectrum even steeper). 
For heavy nuclei, $e Z R_\star$ may be large enough to allow UHECR spectra consistent with the data, but this would mean that they have to dominate in the entire UHECR regime, which is inconsistent with the measured composition. 
We can therefore already get to the preliminary conclusion that the \textit{cosmic-ray spectrum averaged over the radio luminosity function cannot explain the known UHECR data}.
\subsection{Cygnus A}
The UHECR contribution by non-local, ultra-luminous RGs like the FR II source Cygnus A is not contained in the previously introduced average non-local source contribution (\ref{eq:SourceFunctionEvolution}).
Although, Cygnus A with a CR luminosity $Q_{{\rm cr}}(L)=1.3\times 10^{45}\,g_{{\rm cr}}\,\text{erg/s}$ and a corresponding maximal rigidity $\hat{R}(L)=62.6\,g_{{\rm acc}}\,\sqrt{g_{{\rm cr}}}\,\text{EV}$ at a distance of $255\,\text{Mpc}$ \cite{2012A&A...544A..18V}, represents an ideal UHECR accelerator. 
Due to the lack of a reliable EGMF structure out to its distance as well as other technical limitations, we use a 
simplified approach where deflections in the EGMF are neglected. Nevertheless, we are still able to normalize the flux 
in the same manner as we do it for the other sources.  
Deflections by the EGMF are then estimated using a homogeneous random field with a given mean field strength and coherence length. Deflections inside the Galaxy can still be applied in the usual way. 
\section{Best-fit results}
Finally, the previously introduced parameters within their plausibility ranges, i.e. the spectral index $a\in [1.7, 2.2]$, the cosmic-ray load $g_{\rm cr} \in [1, 50]$, the acceleration efficiency $g_{\rm acc} \in [0.01, 0.5]$, and the abundance parameter $q \in [0,2]$, are used to fit the observational data of the flux and mean logarithm of the mass number, $\langle \ln(A) \rangle$, above $5\,\text{EeV}$ from Auger. 
The source-evolution parameter was fixed to $m=3$ as its influence is mainly important below the ankle \cite{1988A&A...199....1B}. 

The normalization of the observed $\langle \ln(A) \rangle$ data depends strongly on the used hadronic interaction model, 
therefore the fitting is performed for the three different contemporary models, QGSJetII-04, Epos-LHC, and Sibyll2.1, 
that have been used by Auger \cite{Aab:2014aea}.

The first fit approaches showed that the UHECR data cannot be explained, if all sources provide the same parameter values. 

\begin{figure}[tbh]
  \centering
    \includegraphics[width=0.49\textwidth]{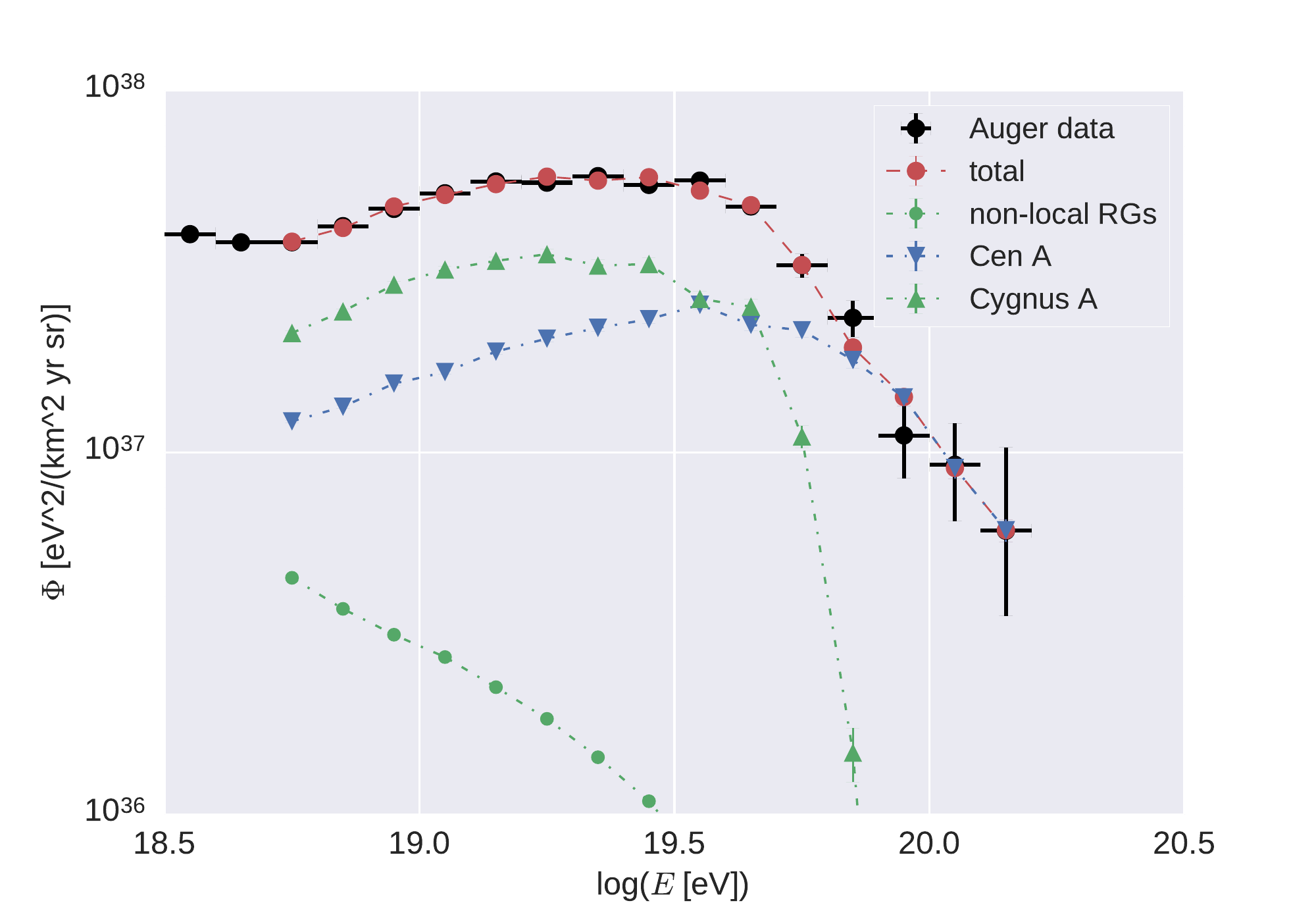}
    \includegraphics[width=0.49\textwidth]{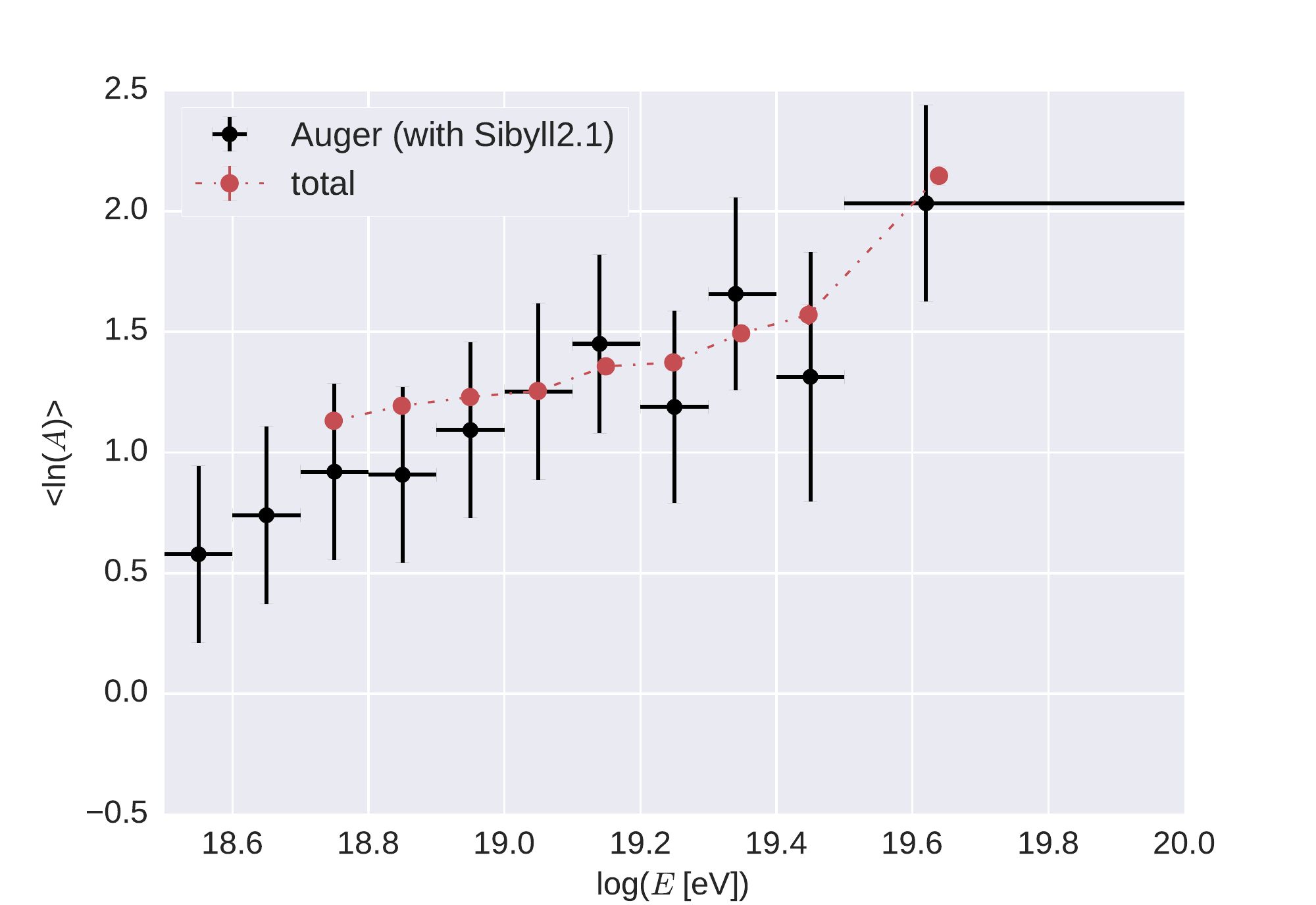}
\caption{Best-fit results to energy spectrum (left) and chemical composition (right) using Sibyll2.1.}
\label{Sibyll_Fit}
\end{figure}

So, we enabled exceptional values for certain parameters of the dominant sources (as summarized in table \ref{FitPar1}) and obtained an accurate explanation of the observational data as shown in Fig.~\ref{Sibyll_Fit}. 
As indicated by the reduced chi-squared value $\chi^2$, the accuracy of the fit does not significantly differ with respect to the used hadronic interaction model. 
The results show that the UHECR data is accurately explained by the CR contributions from Centaurus A and Cygnus A with an isotropically ejected CR power above the ankle of $Q^{\rm CenA}_{\rm uhecr}\simeq 2\times 10^{43}\,\text{erg/s}$ and $Q^{\rm CygA}_{\rm uhecr}\simeq 2\times 10^{47}\,\text{erg/s}$, weakly dependent on the chosen scenario or the hadronic interaction model, while the contribution of all other RGs is irrelevant. 
In particular the average non-local sources are kept negligible by a significantly smaller parameter value of $\bar{g}_{\rm cr}$ compared to the values of Centaurus A or Cygnus A due to its steep spectral behavior. However, an inclusion of some data below the ankle might also increase its significance, and the value of $\bar{g}_{\rm cr}$ could become comparable to the one from Centaurus A and Cygnus A. 

\begin{figure}[tbh]
  \centering
    \includegraphics[width=0.49\textwidth]{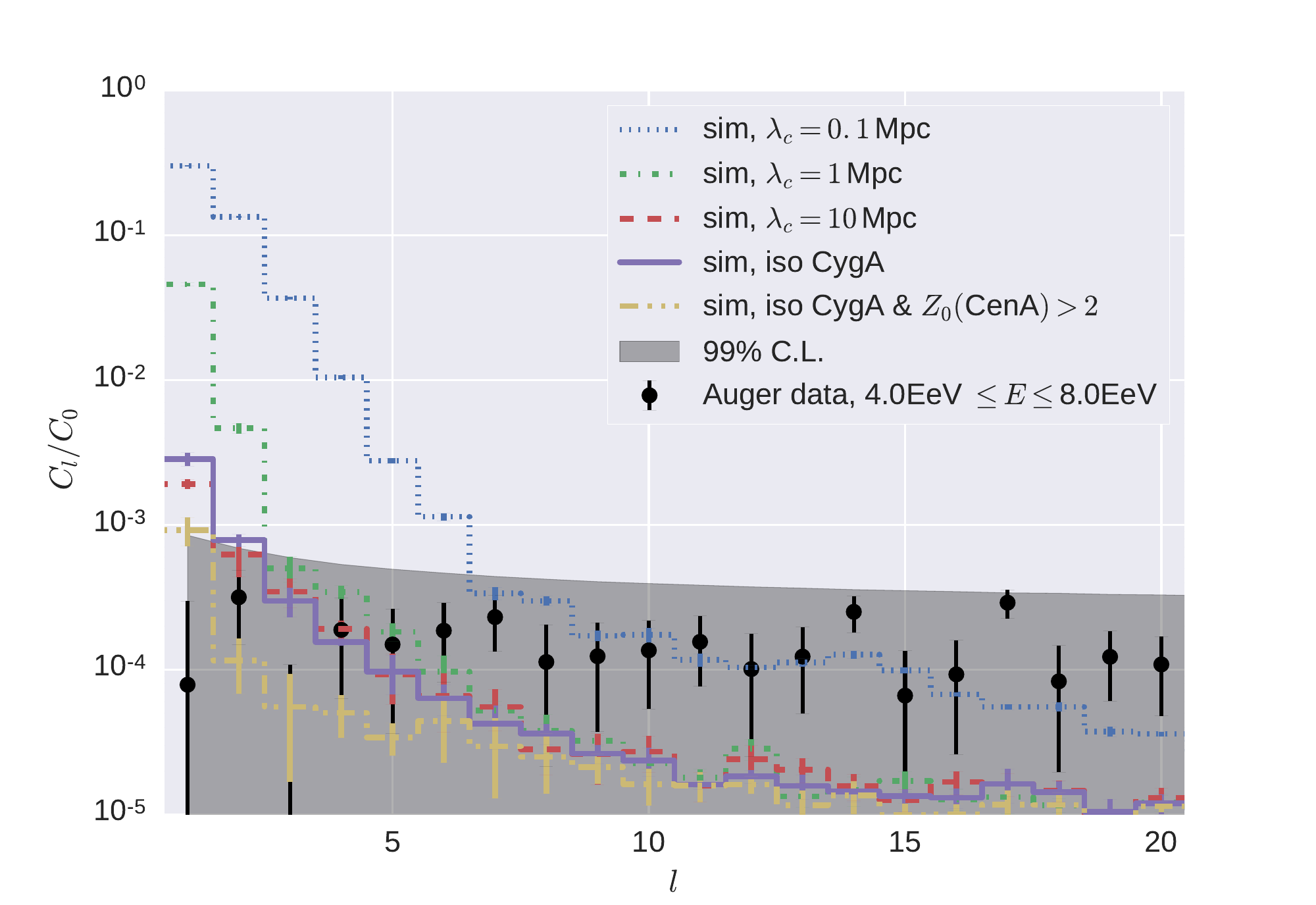}
    \includegraphics[width=0.49\textwidth]{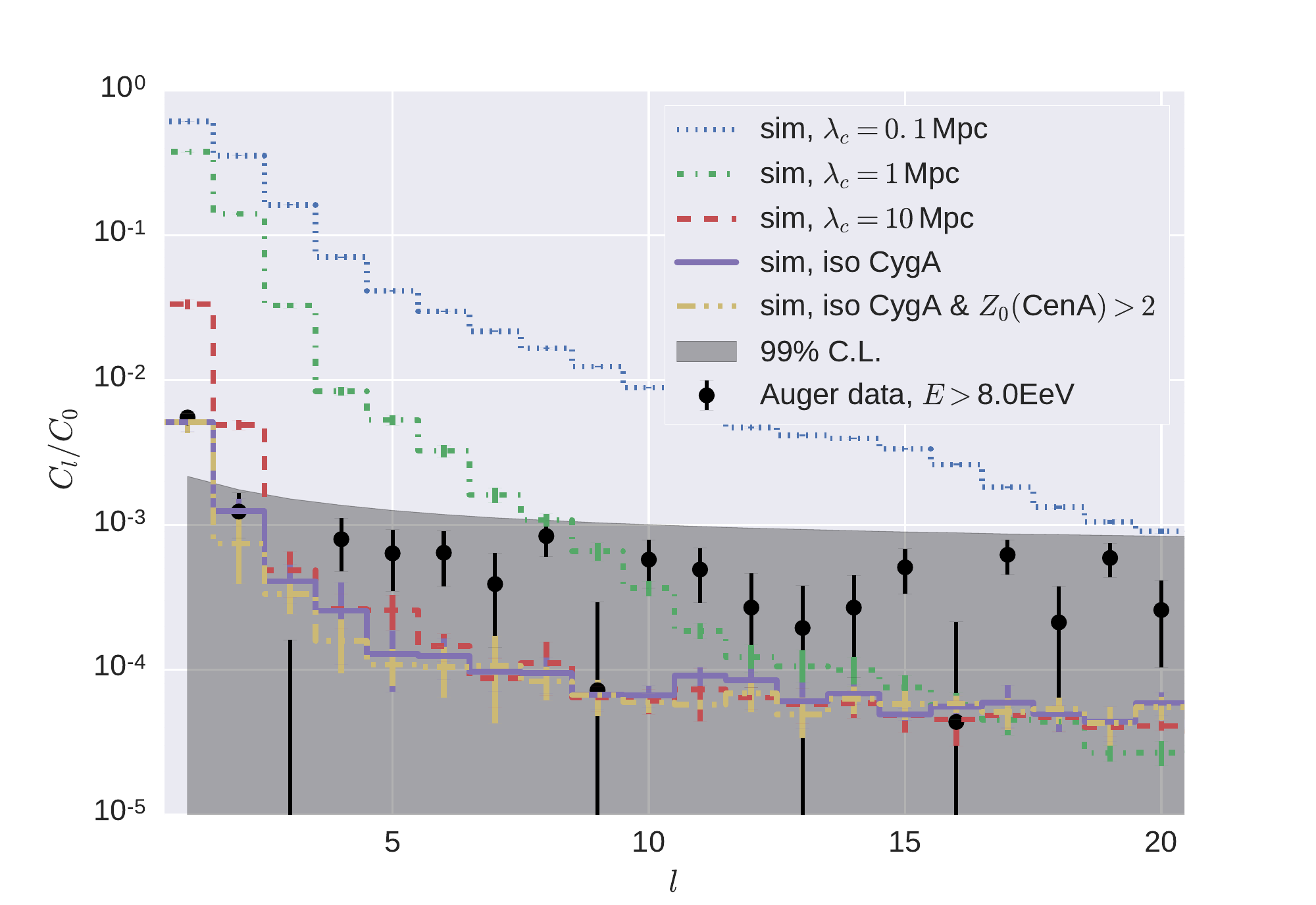}
\caption{Angular power spectrum with isotropized (solid and dash-dotted line) and non-deflected (dashed line) Cygnus A events for $4\,\text{EeV}\leq E \leq 8\,\text{EeV}$ (left), and $E>8\,\text{EeV}$ (right) using best-fit parameters in the case of Sibyll2.1 (see table \ref{FitPar1}).}
\label{Sibyll_PowSpec}
\end{figure}

\begin{table}[h!]
\centering
\caption{Best-fit parameters, with fixed $\bar{q}^{\,\rm CygA}=0$.}
  \begin{tabular}{ l c c c c c c c c}
  \toprule
              & $a$ & $\bar{g}_{\rm cr}$ & $g_{\rm cr}^{\rm CenA}$ & $g_{\rm cr}^{\rm CygA}$ & $\bar{g}^{\rm CenA}_{\rm acc}$ & $g_{\rm acc}^{\rm CygA}$ & $q^{\rm CenA}$ & $\chi^2$  \\ 
  \midrule
    \textbf{EPOS-LHC} & $1.85$ & $7.73$ & $41.54$ & $43.94$ & $0.127$ & $0.059$ & $2$ & $1.1$  \\ 
     \textbf{QGSJetII-04} & $1.82$ & $6.31$ & $21.20$ & $48.84$ & $0.220$ & $0.056$ & $2$ & $1.4$ \\
     \textbf{Sibyll2.1} & $1.83$ & $6.67$ & $24.77$ & $47.90$ & $0.19$ & $0.056$ & $2$ & $1.3$ \\
   \bottomrule
   \multicolumn{9}{l}{\footnotesize Upper index of the parameter indicates the corresponding source (Centaurus A or Cygnus A),}\\
   \multicolumn{9}{l}{\footnotesize bar on top of the parameter corresponds to all sources except Centaurus A and Cygnus A}\\
\end{tabular}
  \label{FitPar1}
\end{table}
To estimate EGMF deflections for Cygnus A, we derived the rms deflection angle for a distance $d=255\,\text{Mpc}$ analytically \cite{2004PhRvD..70d3007S} by
\be
\theta_{k}\simeq 0.8\degree \,\bar{Z}_k\,\left( {\bar{E}_k \over 100\,\text{EeV}} \right)^{-1}\,\left( {d \over 10\,\text{Mpc}} \right)^{1/2}\,\left( {\lambda_{\rm c} \over 1\,\text{Mpc}} \right)^{1/2} \,\left( {B_{\rm rms} \over 1\,\text{nG}} \right)\quad,
\label{rms-deflection}
\ee
where we used an approximated mean candidate energy $\bar{E}_k=(E_k+E_k^*)/2$, an approximated mean candidate charge number $\bar{Z}_k=(Z_k+Z_k^*)/2$, and used as the rms strength $B_{\rm rms}=1.18\,\text{nG}$ of the D05 field. 
Further, a correlation length $0.1\,\text{Mpc}\leq\lambda_{\rm c}\leq 10\,\text{Mpc}$ is assumed, as for $\lambda_{\rm c}\ll 0.1\,\text{Mpc}$ we obtained negligible deflections, whereas in the case of $\lambda_{\rm c}\gg 10\,\text{Mpc}$ the UHECRs with $E_i\lesssim 10\,\text{EeV}$ from Cygnus A were completely isotropized. 
Using the previously derived best-fit parameters the Fig.~\ref{Sibyll_PowSpec} shows the resulting angular power spectra of the arrival directions for different deflection scenarios: Three different coherence lengths $\lambda_{\rm c}$ of a turbulent field with $B_{\rm rms}=1.18\,\text{nG}$, as well as the case of the complete isotropization (i.e.\ $\lambda_{\rm c}\gg 10\,\text{Mpc}$), and the complete isotropization where initial H and He candidates from Centaurus A are excluded. The latter assumption does not affect the total energy spectrum or the chemical composition, but lowers the level of anisotropy and improves consistency with the data set at $4{-}8\,$EeV. 
The motivation for this is that models for a particular heavy enhancement of Centaurus A would expect a pure heavy composition \cite{2010ApJ...720L.155G}. We also note that at $4{-}8\,$ EeV an impact of cosmic rays from the Galaxy or other extragalactic components cannot be excluded, and that at $E > 8\,$EeV our results are in good agreement with the data, for the case that UHECR from Cygnus A are completely isotropized.  
Further, an agreement with the observed arrival directions at energies $E > 8\,$EeV is obtained for 
\be
\label{rms-defl-condition}
\lambda_{\rm c}^{1/2}\, B_{\rm rms} \geq 6\,\text{Mpc}^{1/2}\,\,\text{nG},
\ee
i.e.\ a global coherence length $\lambda_{\rm c}=26\,\text{Mpc}$ in the case of a rms field strength of $B_{\rm rms}\geq 1.18\,\text{nG}$. 
\section{Discussion}
The most solid conclusion that can be drawn is that an average contribution of all radio galaxies cannot explain both 
spectrum and composition of UHECR. 

The second conclusion is, that we need some outstanding, individual sources which explain the cosmic ray flux above the ankle. Here, two sources recommend themselves: Cygnus A, the by far brightest radio galaxy in the sky, and Centaurus A, the second-brightest radio galaxy and also the one nearest to Earth. 
In doing so, either both sources have a cosmic ray load significantly above the average of the bulk of radio galaxies, 
or the average non-local source contribution explains the CRs between $\sim 1\,\text{EeV}$ and the ankle, which seems 
likely due to its spectral behavior.
In addition, it cannot be ruled out that there is also a significant contribution by the other two exceptionally bright local radio galaxies, M87 and Fornax A, but as showed in E+18 they cannot solve the issue alone --- Centaurus A is needed. 

The third conclusion is that Cygnus A needs to provide a solar-like composition, at least if its contribution dominates over the average of the bulk of radio galaxies at the ankle, while Centaurus A needs to have a heavy composition, with an iron fraction comparable to protons at a given cosmic-ray energy. 

A necessary condition for all this to work is that UHECRs from Cygnus A are significantly deflected by the EGMF. We estimate that for $\lambda_{\rm c}^{1/2}\, B_{\rm rms} = 6\,\text{Mpc}^{1/2}\,\,\text{nG}$ all anisotropy constraints at energies ${>}\,8\,$EeV are satisfied. 

Finally, the dominance of a few single sources, whereof Centaurus A is only visible by Auger and Cygnus A is predominantly visible by TA, indicates the need for future investigations that include the impact of the GMF as well as the geometrical exposure of the experiments on the UHECRs from these sources. 
Further, the influence of the EGMF on the energy spectrum and composition of Cygnus A needs to be taken into account and a possible average contribution by the bulk of radio galaxies between the ankle and the second knee needs to be investigated.

\section*{References}
\bibliographystyle{iopart-num}%
\addcontentsline{toc}{section}{Bibliography}
\bibliography{references}

\providecommand{\newblock}{}
\begin{thebibliography}{10}
\expandafter\ifx\csname url\endcsname\relax
  \def\url#1{{\tt #1}}\fi
\expandafter\ifx\csname urlprefix\endcsname\relax\def\urlprefix{URL }\fi
\providecommand{\eprint}[2][]{\url{#2}}

\bibitem{Abbasi:2007sv}
Abbasi R~U {\em et~al.\/} (HiRes) 2008 {\em Phys. Rev. Lett.\/} {\bf 100}
  101101 (\textit{Preprint} \eprint{astro-ph/0703099})

\bibitem{2010PhLB..685..239A}
{Abraham} J {\em et~al.\/} (Pierre Auger) 2010 {\em Phys. Lett. B\/} {\bf 685}
  239--246 (\textit{Preprint} \eprint{1002.1975})

\bibitem{Abraham:2010yv}
Abraham J {\em et~al.\/} (Pierre Auger) 2010 {\em Phys. Rev. Lett.\/} {\bf 104}
  091101 (\textit{Preprint} \eprint{1002.0699})

\bibitem{Aab:2014aea}
Aab A {\em et~al.\/} (Pierre Auger) 2014 {\em Phys. Rev.\/} {\bf D90} 122006
  (\textit{Preprint} \eprint{1409.5083})

\bibitem{ObservatoryMichaelUngerforthePierreAuger:2017fhr}
Unger for~the Pierre Auger~Collaboration M (Pierre Auger) 2017 {\em PoS\/} {\bf
  ICRC2017} 1102 (\textit{Preprint} \eprint{1710.09478})

\bibitem{Aab:2017tyv}
Aab A {\em et~al.\/} (Pierre Auger) 2017 {\em Science\/} {\bf 357} 1266--1270
  (\textit{Preprint} \eprint{1709.07321})

\bibitem{1475-7516-2018-02-036}
Eichmann B, Rachen J, Merten L, van Vliet A and Tjus J~B 2018 {\em Journal of
  Cosmology and Astroparticle Physics\/} {\bf 2018} 036

\bibitem{1475-7516-2016-05-038}
Alves~Batista R {\em et~al.\/} 2016 {\em JCAP\/} {\bf 1605} 038
  (\textit{Preprint} \eprint{1603.07142})

\bibitem{1999MNRAS.309.1017W}
{Willott} C~J, {Rawlings} S, {Blundell} K~M and {Lacy} M 1999 {\em \mnras\/}
  {\bf 309} 1017--1033 (\textit{Preprint} \eprint{astro-ph/9905388})

\bibitem{1970ranp.book.....P}
{Pacholczyk} A~G 1970 {\em {Radio Astrophysics. Nonthermal Processes in
  Galactic and Extragalactic Sources}\/} (W.~H.~Freeman \& Co Ltd, San
  Francisco)

\bibitem{2007MNRAS.375..931M}
{Mauch} T and {Sadler} E~M 2007 {\em \mnras\/} {\bf 375} 931--950
  (\textit{Preprint} \eprint{astro-ph/0612018})

\bibitem{2005JCAP...01..009D}
{Dolag} K, {Grasso} D, {Springel} V and {Tkachev} I 2005 {\em \jcap\/} {\bf 1}
  009 (\textit{Preprint} \eprint{astro-ph/0410419})

\bibitem{2012ApJ...757...14J}
{Jansson} R and {Farrar} G~R 2012 {\em \apj\/} {\bf 757} 14 (\textit{Preprint}
  \eprint{1204.3662})

\bibitem{2012A&A...544A..18V}
{van Velzen} S, {Falcke} H, {Schellart} P, {Nierstenh{\"o}fer} N and {Kampert}
  K~H 2012 {\em \aap\/} {\bf 544} A18 (\textit{Preprint} \eprint{1206.0031})

\bibitem{1988A&A...199....1B}
{Berezinskii} V~S and {Grigor'eva} S~I 1988 {\em \aap\/} {\bf 199} 1--12

\bibitem{2004PhRvD..70d3007S}
{Sigl} G, {Miniati} F and {En{\ss}lin} T~A 2004 {\em \prd\/} {\bf 70} 043007
  (\textit{Preprint} \eprint{astro-ph/0401084})

\bibitem{2010ApJ...720L.155G}
{Gopal-Krishna}, {Biermann} P~L, {de Souza} V and {Wiita} P~J 2010 {\em
  \apjl\/} {\bf 720} L155--L158 (\textit{Preprint} \eprint{1006.5022})

\end{thebibliography}

\end{document}